\documentclass{article}

\usepackage{arxiv}

\usepackage[utf8]{inputenc} 
\usepackage[T1]{fontenc}    
\usepackage{lmodern}        
\usepackage{hyperref}       
\usepackage{url}            
\usepackage{booktabs}       
\usepackage{amsfonts}       
\usepackage{nicefrac}       
\usepackage{microtype}      
\usepackage{graphicx}

\title{Forecasting seasonal criminality using SARIMA\(:\) an application
to monthly aggravated assaults in California}

\author{
    Lucas Hahn
   \\
    Department of Statistics and Applied Probability \\
    University of California, Santa Barbara \\
  Santa Barbara, CA 93117 \\
  \texttt{\href{mailto:lwh152@ucsb.edu}{\nolinkurl{lwh152@ucsb.edu}}} \\
  }

\newlength{\cslhangindent}
\newlength{\csllabelwidth}
\newlength{\cslentryspacingunit} 
  {}%
  {\par}
\newenvironment{CSLReferences}[2] 
 {
  \setlength{\parindent}{0pt}
  \ifodd #1
  \let\oldpar\par
  \def\par{\hangindent=\cslhangindent\oldpar}
  \fi
  \setlength{\parskip}{#2\cslentryspacingunit}
 }%
 {}
\usepackage{calc}

\begin{document}
\maketitle

\begin{abstract}
California experienced an increase in violent criminality during the
last decade, largely driven by a surge in aggravated assaults. To
address this challenge, accurate and timely forecasts of criminal
activity may help state authorities plan ahead and distribute public
resources efficiently to reduce crime. This paper forecasts monthly
aggravated assaults in California using a publicly available dataset on
state crimes and a time series SARIMA model that incorporates the highly
seasonal behavior observed in the data. Results show that predictions
with reasonable accuracy up to six months in advance are produced,
showing the usefulness of these techniques to anticipate state-level
criminal patterns and inform public policy.
\end{abstract}

\keywords{
    California
   \and
    Crime
   \and
    Forecast
   \and
    SARIMA
   \and
    Seasonal
   \and
    Time series
  }

\hypertarget{introduction}{%
\section{Introduction}\label{introduction}}

Criminal activity is widely known to be unevenly distributed over space
and time. It tends to concentrate at particular locations and moments
following patterns that vary depending on the crime. These patterns are
constantly being studied by the literature to better understand their
behavior and inform public policies oriented to reduce criminality. In
this regard, a useful tool for policy makers is the implementation of
statistical models that seek to forecast crime. Producing accurate
forecasts ahead of time may contribute to improve the planning and
deployment of police resources, thus effectively preventing criminal
activity.

In recent years the state of California has experienced an increase in
violent criminality. According to the Criminal Justice Statistics Center
from the Department of Justice of California, violent crimes committed
in the state grew by 2.7\% on average between 2014 and 2019 (Open
Justice n.d.). Violent crimes consist of four different categories:
homicide, rape, robbery and aggravated assault. The increase in
California has been largely driven by a growing number of aggravated
assaults, which went from 91.681 committed offenses in 2014 to 104.756
in 2019. It represents a reversal of the previous decreasing trend
experienced since 1993. The increase in criminality concerns state
residents and visitors alike, whose safety and wellbeing may be
affected.

This paper forecasts monthly aggravated assaults in California using a
publicly available dataset on criminal activity in the state. The data
show strong signs of seasonality, a pattern previously identified by
studies that analyzed violent crimes. Thus, forecasting aggravated
assaults requires the use of a statistical framework that incorporates
seasonality into its structure, a feature that time series models are
well suited to provide. Forecasts up to six months ahead show a
reasonable predictive accuracy, a result that may prove helpful to
inform public policies to prevent criminal activity in California.

The remaining sections of this paper are organized as follows. The next
section presents a review of the relevant literature. Sections 3 and 4
describe the data and the forecasting methodology respectively. Section
5 presents the results of the analysis and Section 6 concludes.

\hypertarget{literature-review}{%
\section{Literature review}\label{literature-review}}

Research on crime forecasting has grown substantially during the last
decades. The development of more accurate predictive methodologies, an
increasing availability of statistical sources and the advancement in
computing capacity are factors that have increased the interest on the
field (Perry 2013). A large body of this literature is related to
spatial analysis of crime and its prediction (Anselin et al. 2000;
Bernasco and Elffers 2010; Gorr and Harries 2003). The spatial
concentration of criminal activity is a feature that has been widely
documented by the literature (Brantingham and Brantingham 1981;
Weisburd, Bernasco, and Bruinsma 2008). In this regard, particular
topics of interest have been the existence and of hot-spots (Eck et al.
2005; Sherman, Gartin, and Buerger 1989) and the interaction between law
enforcement and space (Lee et al. 2010; Lee, Vaughn, and Lim 2014).
Several theories from environmental criminology, such as Crime Pattern
Theory (Brantingham and Brantingham 1984) and Routine Activity Theory
(L. E. Cohen and Felson 1979), provided a theoretical framework to
better understand the observed relation between space and criminal
activity.

In recent years, advances in crime forecasting have been focused on the
implementation of Machine Learning (ML) algorithms because of its
predictive capability. Neural networks, random forests and support
vector machines are some of the techniques considered by the literature
to predict criminal activity (Kounadi et al. 2020; Shamsuddin, Ali, and
Alwee 2017; Hassani et al. 2016). The estimation of a ML model requires
an extensive use of statistical information, which is used to train and
test the performance of the model. For instance, one example of a ML
implementation to predict crime is given by Lin, Yen, and Yu (2018). The
authors predicted vehicle theft using a grid-based approach for the city
of Taoyuan, Taiwan. They compared several ML algorithms and found that a
deep neural network outperformed other models with respect to their
predictive accuracy.

The main advantage of ML algorithms is their capacity to perform well
out of sample. However, their success depends on the availability of
large amounts of data, as well as the correct identification of
appropriate model architectures. For example, recurrent neural networks
and convolutional neural networks are two ML algorithms that differ
substantially in their architecture, and therefore are used for
different predictive tasks (Goodfellow, Bengio, and Courville 2016). In
addition, some ML techniques may have difficulty quantifying the
uncertainty of their predictions (Gawlikowski et al. 2021). This occurs
when the research question requires producing an assessment of
confidence in the predictions of the model, rather than only providing a
point estimate.

When analyzing a single spatial location over time, statistical models
from the field of time series analysis offer a rigorous and reliable
framework (Brockwell and Davis 2009; Shumway, Stoffer, and Stoffer 2000;
Hamilton 2020). They provide flexibility to assimilate empirical
patterns such as seasonality and temporal trends. Moreover, their
theoretical foundations give a structure that allows to efficiently
estimate their models without requiring extensive amounts of data, as is
the case for ML. Time series models have also the advantage of producing
statistical inference and estimates of confidence intervals that allow
to assess the uncertainty of their predictions.

When studying criminality, the incorporation of cyclical patterns into
the modelling structure may be necessary. The empirical literature has
long pointed out that specific types of criminal activity contain a
strong seasonal behavior. For instance, J. Cohen (1941) reviews
different studies for countries in Europe that documented the presence
of summer peaks for violent crimes. Landau and Fridman (1993) studied
robberies and homicides in Israel between 1977 and 1985 and found that
the former reached their peak during winter, while the latter did in
August. With respect to property crimes, Gorr, Olligschlaeger, and
Thompson (2003) mentions that city level crimes of robbery and burglary
tend to be higher during fall and winter. Borowik, Wawrzyniak, and
Cichosz (2018) find that six different types of daily crimes in Poland
between 2013 and 2016 present some type of seasonality, either within a
year or within a week. Thus, a modelling framework to analyze and
forecast criminal activity should consider a methodology that
incorporates this type of statistical behavior into its structure.

Crime forecasts with time series modelling usually follow
Auto-Regressive Integrated Moving-Average models (ARIMA) and their
seasonally adjusted version (SARIMA). For instance, Chen, Yuan, and Shu
(2008) use an ARIMA model to study weekly property crimes in an urban
setting in China. The authors find that the model performed better than
exponential smoothing methods. Cesario, Catlett, and Talia (2016)
forecasted total criminality on a specific area of Chicago using an
ARIMA model estimated on 14 years of weekly data. They show that
one-year-ahead forecasts produced by their model were 80\% accurate on
testing data. Finally, Landau and Fridman (1993) applied a SARIMA model
for the case of Israel to analyze robberies, given the evidence of a
seasonal component in the data.

\hypertarget{data}{%
\section{Data}\label{data}}

The data come from the \textit{Crimes and Clearances} dataset collected
by the Criminal Justice Statistics Center from the California Department
of Justice (Open Justice n.d.). It contains monthly information on
violent crimes (homicide, rape, robberies and aggravated assaults) and
property crimes (property, motor vehicle theft and larceny theft)
committed in the state of California since 1985. The offenses in the
dataset are disaggregated into categories that differ for each crime.
For example, robberies are categorized by both the type of weapon
(firearm, knife, others) and the location of occurrence (highway,
convenience store, bank).

This study focuses on the four categories of aggravated assault present
in the dataset. Each category corresponds to the type of weapon that was
used during the crime: `Firearm', `Knife or cutting instrument', `Other
weapon' and `Hands, fist, feet'. An aggravated assault is defined as a
violent crime consisting of an unlawful attack to a person with the
purpose of inflicting severe bodily injury (Federal Bureau of
Investigation 2017). It is codified under section 245 of California's
Penal Code with sentences that vary depending on the severity of the
conviction and its classification as either a misdemeanor or a felony.

The selected period of analysis corresponds to 15 years of monthly data
between 2005 and 2019. The last year was taken to be before the start of
the Covid19 pandemic, given that it created social disruption and
changes to patterns of criminal activity that are still being currently
analyzed (Ashby 2020; Boman and Gallupe 2020). Due to the unpredictable
nature of the pandemic and its impact on criminal behavior, this study
will focus on pre-pandemic patterns. Annual descriptive statistics for
the four types of aggravated assaults in California between 2005 and
2019 are shown in Table 1.

\begin{center}\includegraphics[width=0.8\linewidth]{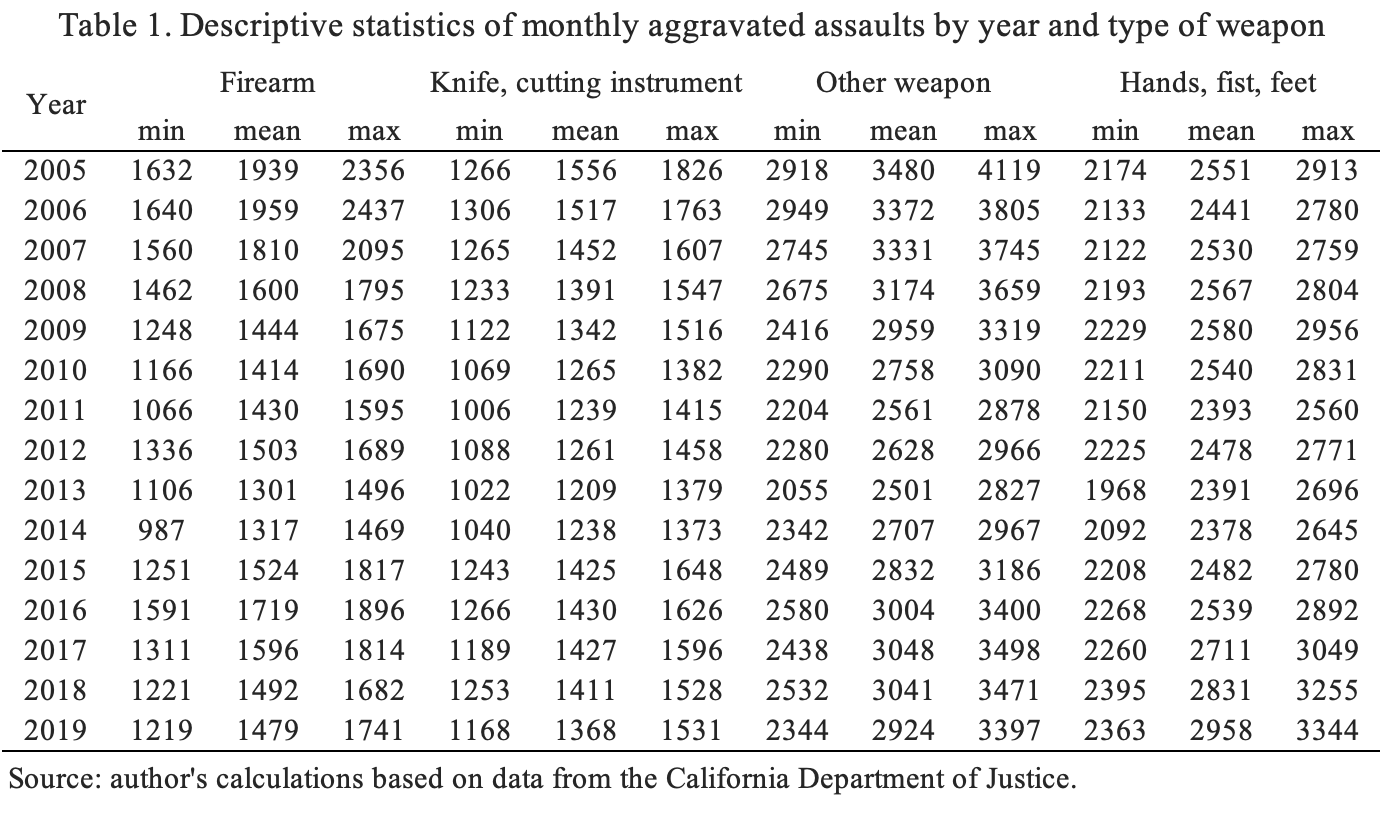} \end{center}

The data on aggravated assaults in California show strong signs of
seasonality. Figure 1 plots the four available categories of monthly
aggravated assaults in the state between 2005 and 2019. The four crimes
peak early in the summer, while the trough is experienced during the
middle of the winter. The seasonal cycle is visually evident on
aggravated assaults that are committed with knifes, other weapons and by
hand. Those committed with firearms also experience the same cyclical
pattern, although with more noise than the previous three categories.
These seasonal patterns are similar in general to those found by the
literature for crimes that have a violent nature (J. Cohen 1941; Landau
and Fridman 1993).

The data also show interesting temporal trends that are worth pointing
out. Aggravated assaults with firearms, knifes and other weapons
experienced a sustained decrease up to 2011, after which they stabilized
for a few years. Those by hand were relatively stable during the same
period of time. After 2014, the four types of assault experienced an
increasing trend that later changed depending on the category. For
firearms, the increasing trend peaked around 2017 and then slowly
decreased until 2019. Knifes and other weapons stabilized between 2016
and 2019, while those by hand continued to increase.

\begin{center}\includegraphics{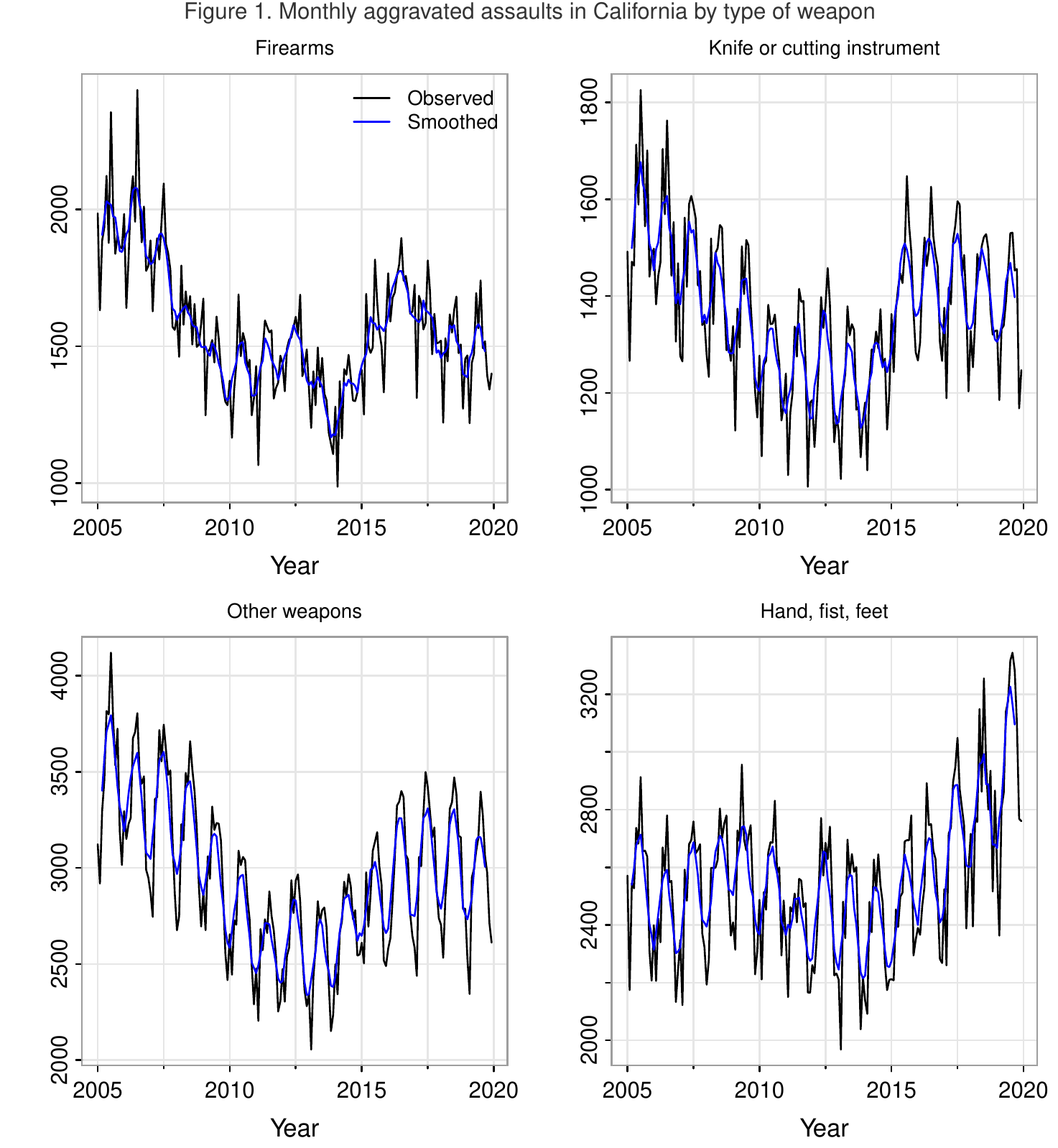} \end{center}

\hypertarget{methodology}{%
\section{Methodology}\label{methodology}}

The presence of a seasonal pattern in the data requires a modelling
framework that incorporates this feature into its structure. Similar to
previous literature, this study proposes a time series SARIMA model to
characterize monthly aggravated assaults in California. SARIMA models
are defined by parameters that shape their seasonal and non-seasonal
components. The procedure to fit the models is as follows. First, each
category of aggravated assaults is differentiated with respect to its
seasonal and non-seasonal lags, in order to obtain a stationary time
series. Then, model parameters that produce the best fit are identified
for each category using the following criteria: i) the autocorrelation
and partial autocorrelation functions; ii) the Akaike Information
Criterion; iii) coefficient significance and the roots of their
corresponding polynomials; and iv) the principle of parsimony, which
states that models with a good fit and fewer parameters should be
preferred.

Additionally, to evaluate the goodness of fit for each model, the
literature recommends performing diagnostic analysis of the model
residuals (Brockwell and Davis 2002). This procedure consists of
checking whether the residual's distribution substantially differs from
a Gaussian distribution, as well as whether they show evidence of
temporal statistical dependence. If any of these conditions is met, then
the model does not provide a proper fit to the data and another
framework should be considered to forecast the data.

The estimation of the models is done using the method of maximum
likelihood. The original data are initially transformed to logarithms in
order to stabilize possible changes in their variance over time. Then,
the data are split into training and testing sets. This guarantees that
model forecasts are compared to data points that were not used during
estimation. The testing set is selected to be the last six months of
data, while the training set corresponds to the remaining previous
points.

\hypertarget{results}{%
\section{Results}\label{results}}

\hypertarget{model-estimation}{%
\subsection{Model estimation}\label{model-estimation}}

An optimal configuration of a SARIMA model is identified for each
category using the previously described criteria. The chosen models are
estimated using R software and their results are presented in the
following table. Model coefficients correspond to non-seasonal
auto-regressive (AR) and moving-average (MA) lags, as well as their
corresponding seasonal versions (SAR, SMA). Standard errors, \(t\)
statistics and their respective p-values are also presented. The
estimations show that every model has at least one statistically
significant seasonal lag. This result reaffirms the importance of
considering methodological frameworks that incorporate seasonal
modelling when analyzing violent criminality.

\begin{center}\includegraphics[width=0.55\linewidth]{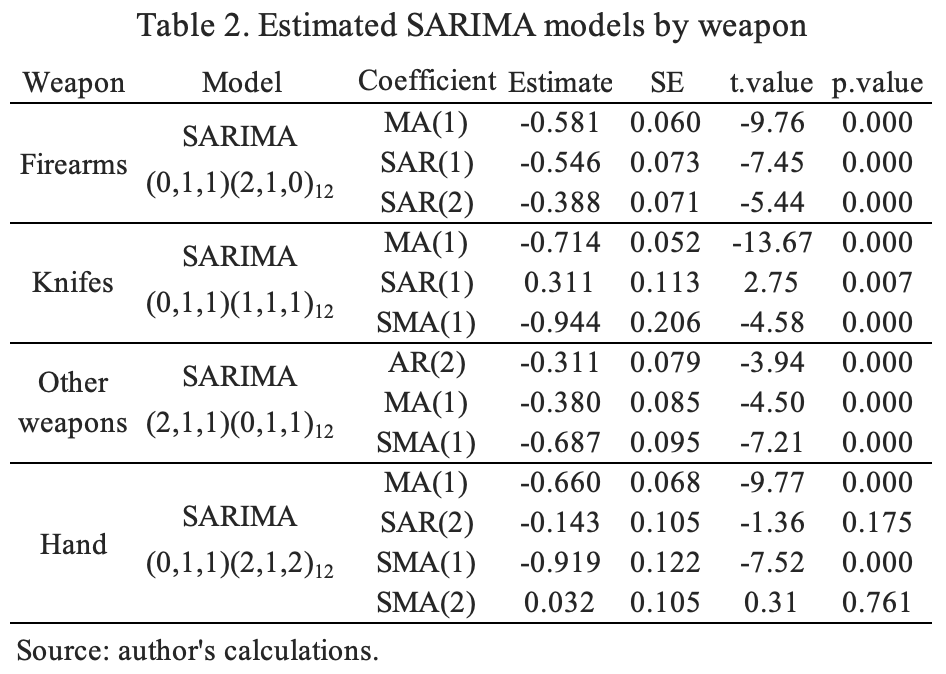} \end{center}

\hypertarget{diagnostic-checking}{%
\subsection{Diagnostic checking}\label{diagnostic-checking}}

To evaluate their goodness of fit, this section performs residual
analysis to each fitted model. Model residuals should resemble a
Gaussian distribution, as well as show no signs of statistical
correlation (Brockwell and Davis 2002). The diagnostic checking is
performed in two steps. The first step is an analysis of the residuals'
distribution. Figure 2 shows the histogram and QQ plot for each category
of crime. Notice that the four models present a good fit to the data,
given that the histograms are symmetric with their modes close to zero.
Moreover, the QQ plots lie close to the identity line, which shows that
the residuals have similar quantiles to those of a Gaussian
distribution.

\begin{center}\includegraphics{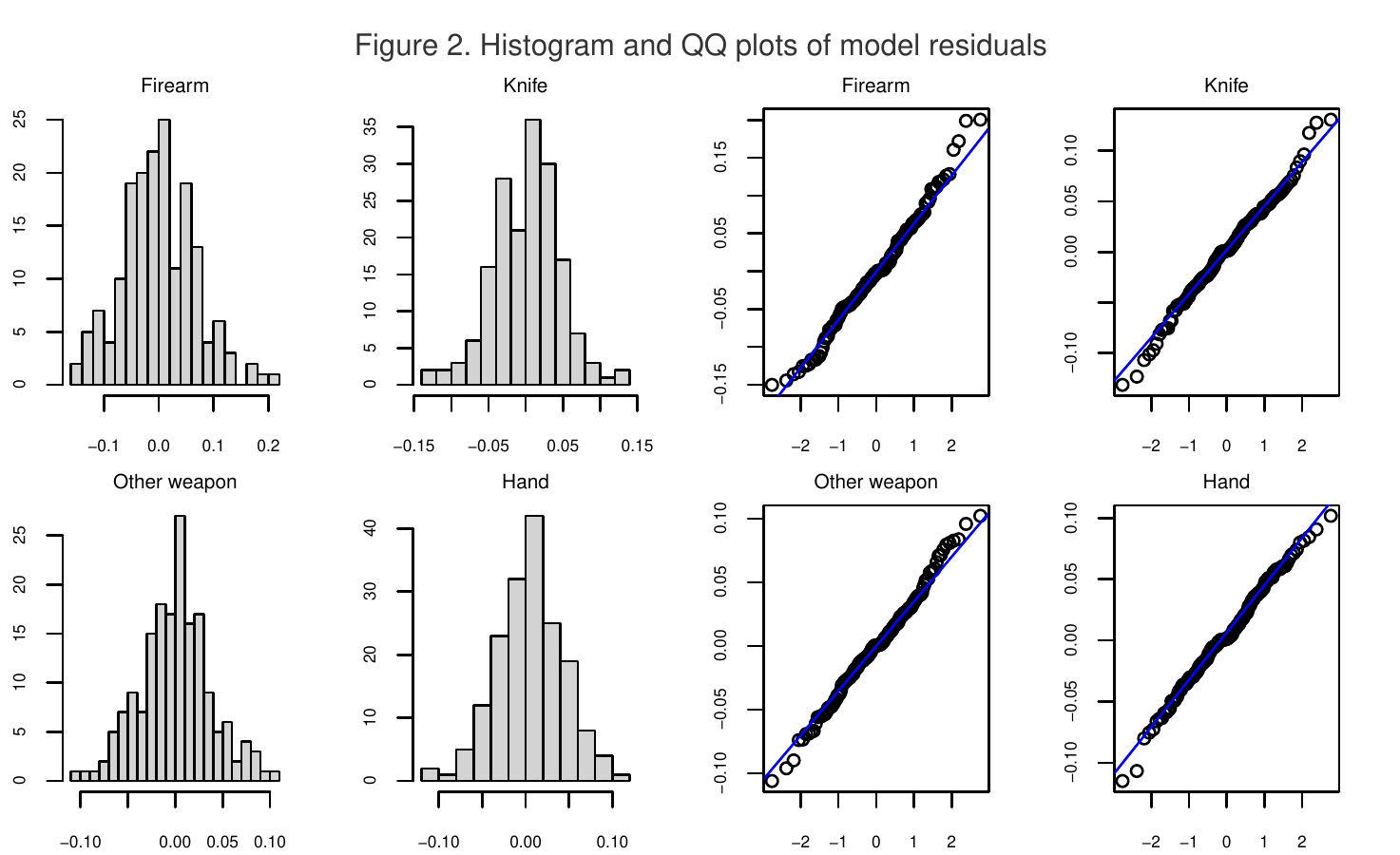} \end{center}

The second step of diagnostic checking consists on the application of
statistical tests for correlation and normality. Correlation is tested
using the Ljung-Box (Ljung and Box 1978) and Box-Pierce (Box and Pierce
1970) tests, while normality is tested using the Shapiro-Wilk test
(Shapiro and Wilk 1965). Table 2 presents the results of the three tests
applied to the four models. The null hypotheses of no correlation and
normality of the residuals are not rejected with a level of 10\% for any
of the four models. Thus, the tests confirm the result from the previous
step that the suggested models provide a good fit to the data.

\begin{center}\includegraphics[width=0.35\linewidth]{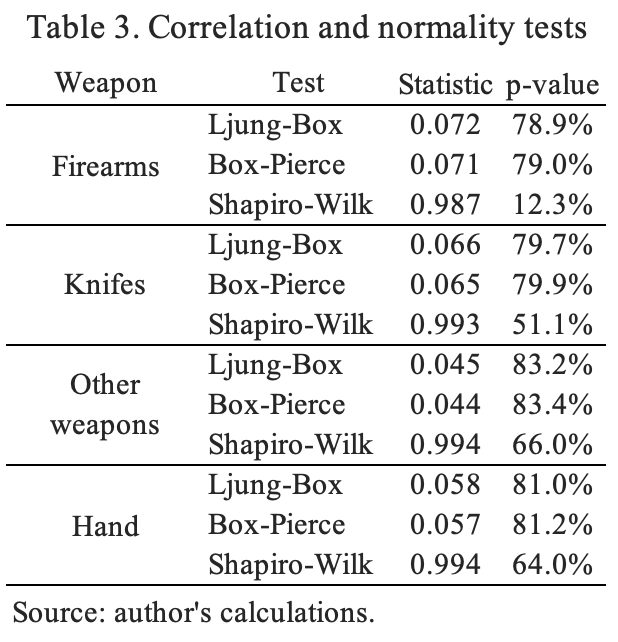} \end{center}

\hypertarget{forecasts}{%
\subsection{Forecasts}\label{forecasts}}

This section forecasts each category of aggravated assaults over a
period of six months. The forecasts and their 90\% confidence intervals
are presented in Figure 3 with the corresponding observed values from
testing data. In all cases the models correctly predicted the negative
trend observed in aggravated assaults that is experienced during the
second half of the year, when violent criminality tends to decrease
after reaching its peak during summer. Additionally, notice that
observed values were mostly contained inside the 90\% confidence regions
predicted by the models, which is an indication of good predictive
accuracy. Moreover, criminal activity six months ahead was forecast
considerably close to the observed values. This capacity to predict up
to six months ahead the level of criminality that the state will
experience can be a valuable tool for state authorities.

A more detailed inspection shows that aggravated assaults with other
weapons presents a highly accurate fit, with forecasts lying very close
to the respective observed values. This particular category shows the
most stable seasonal behavior, thus allowing more accurate forecasts and
narrower confidence intervals. Aggravated assaults with knifes or by
hand are statistically noisier and therefore show wider confidence
intervals. For the latter, there is a minor increase on the observed
values at the beginning of the forecasted period that was not
anticipated by the model. This shows the importance of regularly
updating the estimated models using the most recently available data, so
that the latest trends and seasonal patterns are appropriately
incorporated. Finally, aggravated assaults with firearms show the most
noisy data and thus produce predictions that have the widest confidence
intervals. However, notice that despite the noise the forecasts lie
close to the observed values, showing that it can also be predicted with
reasonable accuracy.

\begin{center}\includegraphics{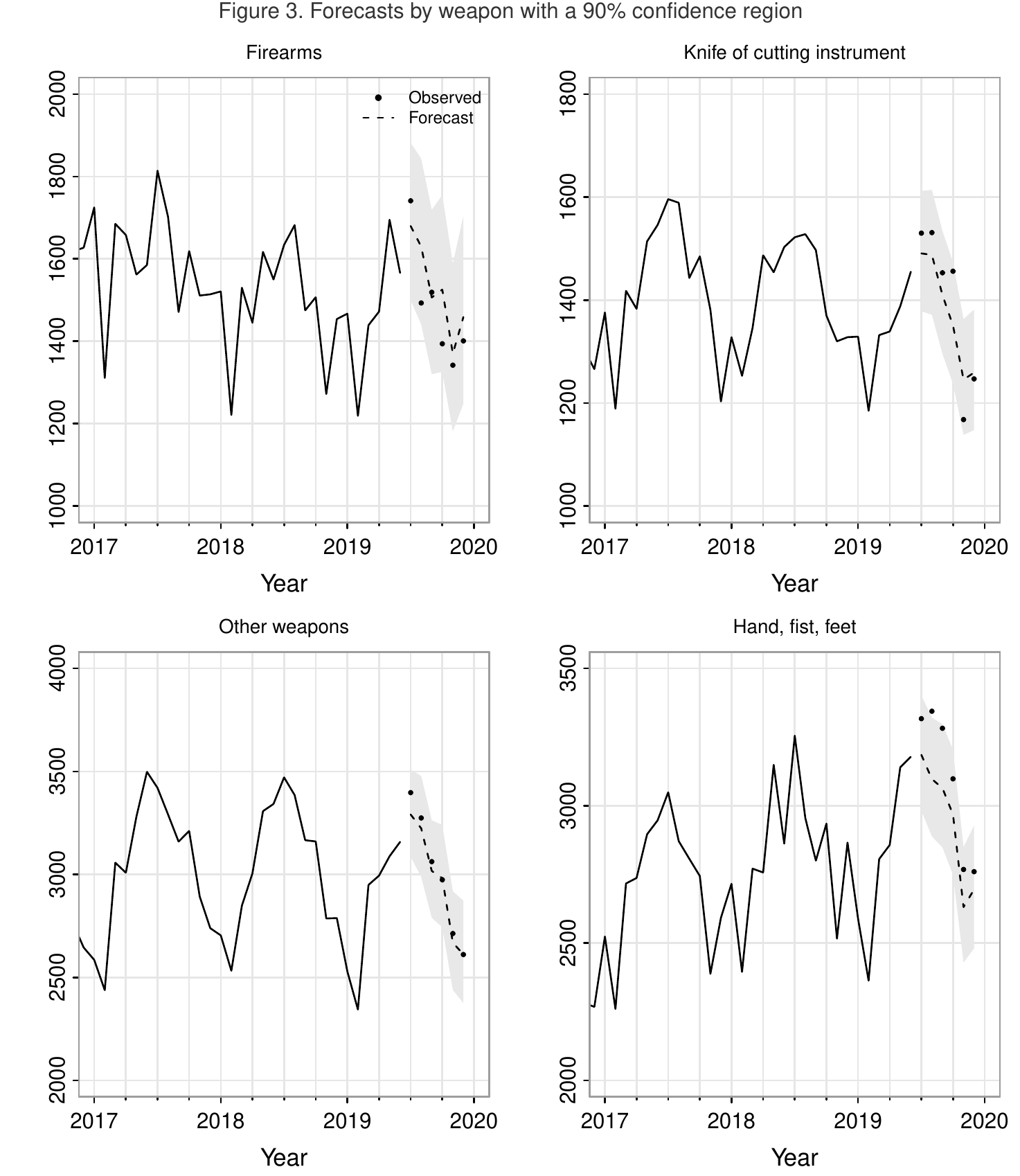} \end{center}

The numerical values of crimes and forecasts with their corresponding
confidence intervals and relative errors are presented in the following
table. All forecasts lie within a 10\% margin of the observed values and
most of them within 5\%. The category of aggravated assault that
presents the higher accuracy is other weapons, with relative errors
between 0\% and 3.1\%. Firearms shows the highest errors with values of
9.4\% and 9.2\% for the months of October and August, respectively. If
only the first monthly forecasts are considered, the highest observed
relative error was 4\%. This is a reasonable predictive accuracy for one
month ahead forecasts. It is also important to notice that statistical
models may be updated recurrently as more recent data becomes available,
thus constantly improving and correcting the forecasts using the latest
trends in the data.

\begin{center}\includegraphics[width=0.55\linewidth]{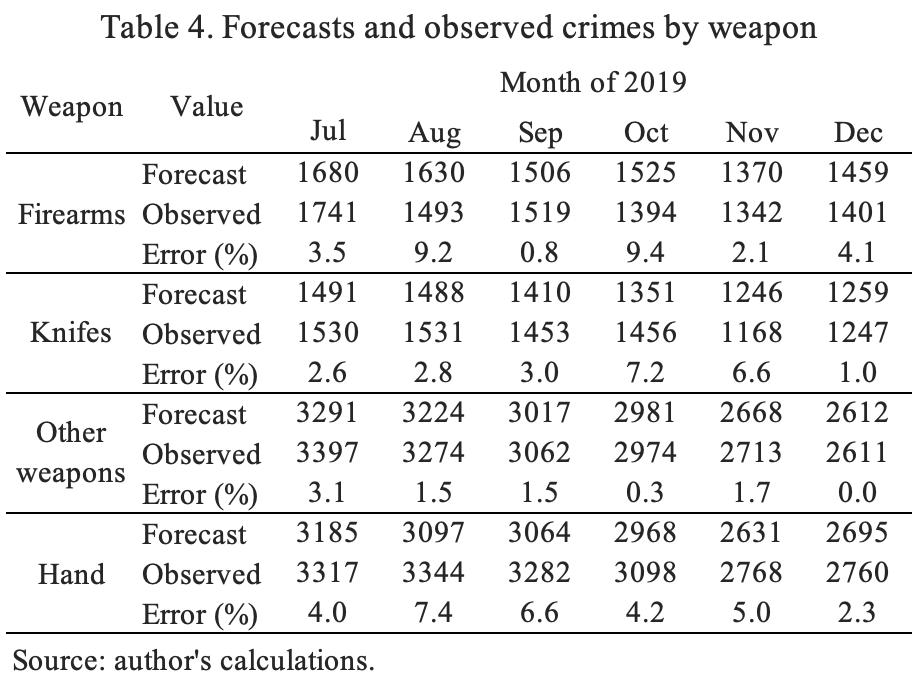} \end{center}

\hypertarget{conclusion}{%
\section{Conclusion}\label{conclusion}}

This paper presented a time series framework to forecast monthly
aggravated assaults in California, a criminal activity with strong
seasonal patterns that experienced an increasing trend over the last
years. A SARIMA model was fit to four different categories of aggravated
assaults and showed reasonable predictive accuracy up to six months in
advance. Accurate and timely forecasts may help public authorities plan
ahead of time and execute better public policies to prevent crime from
occurring. Future work should address the causes behind the increasing
violent criminality observed in California, as well as comparing its
experience to other states and the public policies put in place to
effectively prevent further increases.

\newpage

\hypertarget{references}{%
\section*{References}\label{references}}
\addcontentsline{toc}{section}{References}

\hypertarget{refs}{}
\begin{CSLReferences}{1}{0}
\leavevmode\vadjust pre{\hypertarget{ref-anselin2000spatial}{}}%
Anselin, Luc, Jacqueline Cohen, David Cook, Wilpen Gorr, and George
Tita. 2000. {``Spatial Analyses of Crime.''} \emph{Criminal Justice} 4
(2): 213--62.

\leavevmode\vadjust pre{\hypertarget{ref-ashby2020initial}{}}%
Ashby, Matthew PJ. 2020. {``Initial Evidence on the Relationship Between
the Coronavirus Pandemic and Crime in the United States.''} \emph{Crime
Science} 9 (1): 6.

\leavevmode\vadjust pre{\hypertarget{ref-bernasco2010statistical}{}}%
Bernasco, Wim, and Henk Elffers. 2010. {``Statistical Analysis of
Spatial Crime Data.''} \emph{Handbook of Quantitative Criminology},
699--724.

\leavevmode\vadjust pre{\hypertarget{ref-boman2020has}{}}%
Boman, John H, and Owen Gallupe. 2020. {``Has COVID-19 Changed Crime?
Crime Rates in the United States During the Pandemic.''} \emph{American
Journal of Criminal Justice} 45: 537--45.

\leavevmode\vadjust pre{\hypertarget{ref-borowik2018time}{}}%
Borowik, Grzegorz, Zbigniew M Wawrzyniak, and Paweł Cichosz. 2018.
{``Time Series Analysis for Crime Forecasting.''} In \emph{2018 26th
International Conference on Systems Engineering (ICSEng)}, 1--10. IEEE.

\leavevmode\vadjust pre{\hypertarget{ref-box1970distribution}{}}%
Box, George EP, and David A Pierce. 1970. {``Distribution of Residual
Autocorrelations in Autoregressive-Integrated Moving Average Time Series
Models.''} \emph{Journal of the American Statistical Association} 65
(332): 1509--26.

\leavevmode\vadjust pre{\hypertarget{ref-brantingham1981environmental}{}}%
Brantingham, Paul J, and Patricia L Brantingham. 1981.
\emph{Environmental Criminology}. SAGE Publications, Incorporated.

\leavevmode\vadjust pre{\hypertarget{ref-brantingham1984patterns}{}}%
---------. 1984. \emph{Patterns in Crime}. Macmillan New York.

\leavevmode\vadjust pre{\hypertarget{ref-brockwell2002introduction}{}}%
Brockwell, Peter J, and Richard A Davis. 2002. \emph{Introduction to
Time Series and Forecasting}. Springer.

\leavevmode\vadjust pre{\hypertarget{ref-brockwell2009time}{}}%
---------. 2009. \emph{Time Series: Theory and Methods}. Springer
science \& business media.

\leavevmode\vadjust pre{\hypertarget{ref-cesario2016forecasting}{}}%
Cesario, Eugenio, Charlie Catlett, and Domenico Talia. 2016.
{``Forecasting Crimes Using Autoregressive Models.''} In \emph{2016 IEEE
14th Intl Conf on Dependable, Autonomic and Secure Computing, 14th Intl
Conf on Pervasive Intelligence and Computing, 2nd Intl Conf on Big Data
Intelligence and Computing and Cyber Science and Technology Congress
(DASC/PiCom/DataCom/CyberSciTech)}, 795--802. IEEE.

\leavevmode\vadjust pre{\hypertarget{ref-chen2008forecasting}{}}%
Chen, Peng, Hongyong Yuan, and Xueming Shu. 2008. {``Forecasting Crime
Using the Arima Model.''} In \emph{2008 Fifth International Conference
on Fuzzy Systems and Knowledge Discovery}, 5:627--30. IEEE.

\leavevmode\vadjust pre{\hypertarget{ref-cohen1941geography}{}}%
Cohen, Joseph. 1941. {``The Geography of Crime.''} \emph{The Annals of
the American Academy of Political and Social Science} 217 (1): 29--37.

\leavevmode\vadjust pre{\hypertarget{ref-cohen1979social}{}}%
Cohen, Lawrence E, and Marcus Felson. 1979. {``Social Change and Crime
Rate Trends: A Routine Activity Approach.''} \emph{American Sociological
Review}, 588--608.

\leavevmode\vadjust pre{\hypertarget{ref-eck2005mapping}{}}%
Eck, John, Spencer Chainey, James Cameron, and Ronald Wilson. 2005.
{``Mapping Crime: Understanding Hotspots.''}

\leavevmode\vadjust pre{\hypertarget{ref-fbi}{}}%
Federal Bureau of Investigation. 2017. {``Uniform Crime Reporting
Program: Aggravated Assaults.''}

\leavevmode\vadjust pre{\hypertarget{ref-gawlikowski2021survey}{}}%
Gawlikowski, Jakob, Cedrique Rovile Njieutcheu Tassi, Mohsin Ali,
Jongseok Lee, Matthias Humt, Jianxiang Feng, Anna Kruspe, et al. 2021.
{``A Survey of Uncertainty in Deep Neural Networks.''} \emph{arXiv
Preprint arXiv:2107.03342}.

\leavevmode\vadjust pre{\hypertarget{ref-goodfellow2016deep}{}}%
Goodfellow, Ian, Yoshua Bengio, and Aaron Courville. 2016. \emph{Deep
Learning}. MIT press.

\leavevmode\vadjust pre{\hypertarget{ref-gorr2003introduction}{}}%
Gorr, Wilpen, and Richard Harries. 2003. {``Introduction to Crime
Forecasting.''} \emph{International Journal of Forecasting} 19 (4):
551--55.

\leavevmode\vadjust pre{\hypertarget{ref-gorr2003short}{}}%
Gorr, Wilpen, Andreas Olligschlaeger, and Yvonne Thompson. 2003.
{``Short-Term Forecasting of Crime.''} \emph{International Journal of
Forecasting} 19 (4): 579--94.

\leavevmode\vadjust pre{\hypertarget{ref-hamilton2020time}{}}%
Hamilton, James Douglas. 2020. \emph{Time Series Analysis}. Princeton
University Press.

\leavevmode\vadjust pre{\hypertarget{ref-hassani2016review}{}}%
Hassani, Hossein, Xu Huang, Emmanuel S Silva, and Mansi Ghodsi. 2016.
{``A Review of Data Mining Applications in Crime.''} \emph{Statistical
Analysis and Data Mining: The ASA Data Science Journal} 9 (3): 139--54.

\leavevmode\vadjust pre{\hypertarget{ref-kounadi2020systematic}{}}%
Kounadi, Ourania, Alina Ristea, Adelson Araujo, and Michael Leitner.
2020. {``A Systematic Review on Spatial Crime Forecasting.''}
\emph{Crime Science} 9: 1--22.

\leavevmode\vadjust pre{\hypertarget{ref-landau1993seasonality}{}}%
Landau, Simha F, and Daniel Fridman. 1993. {``The Seasonality of Violent
Crime: The Case of Robbery and Homicide in Israel.''} \emph{Journal of
Research in Crime and Delinquency} 30 (2): 163--91.

\leavevmode\vadjust pre{\hypertarget{ref-lee2010examination}{}}%
Lee, Hoon, Hyunseok Jang, Ilhong Yun, Hyeyoung Lim, and David W Tushaus.
2010. {``An Examination of Police Use of Force Utilizing Police Training
and Neighborhood Contextual Factors: A Multilevel Analysis.''}
\emph{Policing: An International Journal of Police Strategies \&
Management} 33 (4): 681--702.

\leavevmode\vadjust pre{\hypertarget{ref-lee2014impact}{}}%
Lee, Hoon, Michael S Vaughn, and Hyeyoung Lim. 2014. {``The Impact of
Neighborhood Crime Levels on Police Use of Force: An Examination at
Micro and Meso Levels.''} \emph{Journal of Criminal Justice} 42 (6):
491--99.

\leavevmode\vadjust pre{\hypertarget{ref-lin2018grid}{}}%
Lin, Ying-Lung, Meng-Feng Yen, and Liang-Chih Yu. 2018. {``Grid-Based
Crime Prediction Using Geographical Features.''} \emph{ISPRS
International Journal of Geo-Information} 7 (8): 298.

\leavevmode\vadjust pre{\hypertarget{ref-ljung1978measure}{}}%
Ljung, Greta M, and George EP Box. 1978. {``On a Measure of Lack of Fit
in Time Series Models.''} \emph{Biometrika} 65 (2): 297--303.

\leavevmode\vadjust pre{\hypertarget{ref-dojCA}{}}%
Open Justice. n.d. {``Crimes and Clearances - Monthly Data.''}

\leavevmode\vadjust pre{\hypertarget{ref-perry2013predictive}{}}%
Perry, Walt L. 2013. \emph{Predictive Policing: The Role of Crime
Forecasting in Law Enforcement Operations}. Rand Corporation.

\leavevmode\vadjust pre{\hypertarget{ref-8075335}{}}%
Shamsuddin, Nurul Hazwani Mohd, Nor Azizah Ali, and Razana Alwee. 2017.
{``An Overview on Crime Prediction Methods.''} In \emph{2017 6th ICT
International Student Project Conference (ICT-ISPC)}, 1--5.
\url{https://doi.org/10.1109/ICT-ISPC.2017.8075335}.

\leavevmode\vadjust pre{\hypertarget{ref-shapiro1965analysis}{}}%
Shapiro, Samuel Sanford, and Martin B Wilk. 1965. {``An Analysis of
Variance Test for Normality (Complete Samples).''} \emph{Biometrika} 52
(3/4): 591--611.

\leavevmode\vadjust pre{\hypertarget{ref-sherman1989hot}{}}%
Sherman, Lawrence W, Patrick R Gartin, and Michael E Buerger. 1989.
{``Hot Spots of Predatory Crime: Routine Activities and the Criminology
of Place.''} \emph{Criminology} 27 (1): 27--56.

\leavevmode\vadjust pre{\hypertarget{ref-shumway2000time}{}}%
Shumway, Robert H, David S Stoffer, and David S Stoffer. 2000.
\emph{Time Series Analysis and Its Applications}. Vol. 3. Springer.

\leavevmode\vadjust pre{\hypertarget{ref-weisburd2008putting}{}}%
Weisburd, David, Wim Bernasco, and Gerben Bruinsma. 2008. \emph{Putting
Crime in Its Place}. Springer.

\end{CSLReferences}

\end{document}